# Success Factors Contributing to eGovernment Adoption in Saudi Arabia

Ibrahim Khalil Abu Nadi
Dr Louis Sanzogni
Dr Kuldeep Sandhu
Dr Peter Woods
*Griffith University Management Department*
*i.abunadi@griffith.edu.au*

## Abstract

*Saudi Arabia is predetermined to implement eGovernment and provide world-class government services to citizens by 2010. However, this initiative will be meaningless if the people didn't adopt these electronic services.*

*Therefore, the purpose of this study is to determine success factors that will facilitate the adoption of eGovernment in Saudi Arabia. The results of the literature review have been deployed into surveys with Saudi eGovernment users. The discussion of the analysis from results obtained from the practical study has provided a framework that encompasses the eGovernment adoption success factors for Saudi Arabia.*

## 1. Introduction

The KSA (Kingdom of Saudi Arabia) began its own eGovernment project in 2005 with a planned implementation in most of the government agencies by 2010 [20] thus fulfilling the KSA government's future objectives to allow their clients to acquire online government services [5]. The vision statement for the KSA's eGovernment project highlights this goal of user-centricity [20]: "By the end of 2010, everyone in the Kingdom will be able to enjoy from anywhere and at any time- world class government services offered in a seamless, user friendly and secure way by utilizing a variety of electronic means". Despite this aspiring statement, it is not clear whether people will embrace these initiatives [8]. Reasons for uncertainty lie in the fact that the current ICT (Information Communication Technology) infrastructure is still under continual improvement [2]. While the internet growth rate is 1170.0 %, the current internet usage is considerably low 10.7% [17]. Further to this, there are still unanswered questions (with no evidence in any literature, academic or otherwise) on the KSA citizen's attitude toward eGovernment implementation and its usage; their reaction to the proposed and eventual eGovernment implementation and issues of perception in terms of the accessibility to the eGovernment systems.

The purpose of this research therefore, is to investigate the phenomena of eGovernment adoption within the context of the KSA environment. Current and potential users of Saudi eGovernment services will be surveyed.

## 2. EGovernment Adoption Success Factors in the Literature

EGovernment adoption becomes particularly significant owing to the effects of factors such as cost reduction and improving quality of services [22]. A wide number of researches ([15], [2], [11], [16], [6]) have discussed eGovernment adoption success factors within organizational contexts. However, these studies and others ([22], [21], [18]) include various factors that can be applied at the individuals' level [22] discussed a strategy that would encourage eGovernment adoption within a society by building trust. The government should seek the first adoptees that are called transformers [1] and encourage them to use eGovernment. By the transformer's their word-of-mouth eGovernment can have a foundation for citizen-to-citizen advertisement campaign [22]. Joshi et al. (2002) and Warkentin et al. (2002) agree that trust is another major factor for eGovernment adoption. They also agreed with Carter and Belanger (2005) that to increase citizens' trust of eGovernment, institutional measures and guarantees on security and privacy of eGovernment need to be created at least at the C2G (Citizen to Government) level. Security and privacy of eGovernment are important ICT infrastructure components that the government need to adequately implement ([15], [16]).

Reffat (2003) recommends maintaining an ICT infrastructure that is accessible by the country's residence using different technologies such as wireless internet connections, computer centers and kiosks. ICT infrastructure requires updating not only G2C levels [13]. According to Layne and Lee (2001), government agencies and departments require presence of internet communications to be able to provide, update and maintain eGovernment. The integration between eGovernment and ICT technologies usage raises legal and policy issues that require addressing by government policymakers [21] such as individual's identity authenticity on the internet, sending electronic document via ICT technologies or stakeholders' rights in the internet.

Since eGovernment depends on IT, users need some level of IT awareness and proficiency that enables the usage of eGovernment. Reffat (2003) recommends that the Saudi government should provide training campaigns for citizens that will assist those who find difficulty in using technology. Nevertheless, Chen et al. (2006) points out that, citizens of developed countries are more likely to adopt eGovernment than their counterparts in developing countries due to computer literacy provided in schools. Therefore, providing updated computer curriculum in an early stage of schools can contribute to eGovernment success in the long run [10].

Education and training can be important [10]; however it is also important to conduct marketing campaigns to improve the people's awareness of e-services provided [12]. As an example, Australian government surveyed 5040 citizens to find out why there was a lack of eGovernment usage. The highest per cent (32) of the citizen's reason was that they were not aware that the government provides these services online [3]. As a solution Geray and Al Bastaki (2005) recommended conducting mass marketing campaigns online and offline defining and describing the provided online services.

EGovernment websites are representative of the government itself and carefully supporting them affects people's adoption and acceptance. Ease of use is a factor that is discussed in the literature because of its influence on users' adoption [22]. However, other researchers ([14], [22]) state that, ease of use does not affect eGovernment adoption. Between those two arguments perceptions on eGovernment ease of use are still unclear in KSA and require further investigation. Perceived usefulness of eGovernment websites is another factor that can be attached to ease of use. Carter and Belanger (2004) found out that the higher degrees of adoption occur when citizens highly perceive eGovernment usefulness.

Another factor concluded by Carter and Belanger (2003) is using tangible verification to enhance perceived compatibility. In traditional methods people are accustomed to verify their transactions using conventional methods e.g. papers receipts. They suggested when a user launches an online transaction, a paper receipt can be sent to the user upon request through mail or fax [7].

## 3. Methodology

Participants were listed first based on their intention to benefit from the KSA government services. Total number of participants was 316. Some participants were reached using emails (15.8%) others were reached by personal interviews (20.1%) and the remaining were reached using mail (64.1%). The results were analyzed using normalized frequency distribution analysis.

### 3.1 Sample

The number of participants living in urban areas (87.7%) was more than rural (12.3%). Most of the participants were aged between 26 to 35 years old and as for education: 59.3 per cent had a graduate degree, 20.8 per cent had post-graduate degrees and the rest had either technical (5.7%) or high school degree (13.9). Most of the participants (89.9%) were inclined toward the usage of technology and most of them (78.5%) were predisposed to technology in their education or occupation. Saudis (79%.2) comprised most number of participants in this survey; however, there were 13.6 per cent KSA residence and 6.9% visitors.

### 3.2 Variables

The aim of the survey was to test the applicability of previously discussed .The questionnaire was developed with regards the success factors that were summarized from the literature. These aspects were formulated into variables (shown in Table 1 below), used to address the research question.

**Table 1:** Base and dependant variables

| Base Variables | Dependent Variables |
|---|---|
| -Living region (Binary) | ICT accessibility: ICT availability (Binary), ICT cost and quality (Composite). |
| -Income (Composite) -Age (Composite) -Education (Composite) | ICT usage: Internet experience (Binary), Internet years of experience (Composite). |
| -Nationality [Saudi Citizen, Saudi residence, KSA visitor] -ECommerce experience (Binary) | Believes about eGov: Security (Composite), Privacy (Composite) |
| -Disposition toward technology (Binary) -Disposition toward technology in education (Binary) | Behavior intentions to adopt eGov: Ease of use(Composite), Usefulness (Composite) |
| -EGovernment experience (Binary) -ECommerce experience (Binary) | Decision on eGov adoption: Reliability (Composite) |
| -Number of participants | EGov adoption: Perceived best method to contact the government(Composite), best perceived method to increase eGov awareness(Composite), preference of receiving or not a hardcopy evidence of e-transaction status (Binary) |

Table 1 differentiates between base and dependant variables and composite and binary variables. Base variables are aspects that can have effect on dependant variables related to eGovernment adoption. Dependent variables have been grouped and recognized. There were two types of variables binary and composite: binary variables can have two values e.g. living region can be either "urban" or "rural". On the other hand, composite variables can have more than two values e.g. four values has been used for the variable Education in this research, high school, technical qualification, graduate or post-graduate.

## 3.2 Normalized Frequency Distribution Analysis

The normalized frequency distribution method has been used to understand and measure the importance of the relationships between base variables and dependent variables. Additionally it is to test and compare (because of the sequential nature of this research).The analysis will be divided into five phases that are based on a modified model of theory of planned behavior [4] that suits the purposes of this research. The phases will be ordered as following: ICT accessibility, ICT usage, eGovernment awareness, perceptions about eGovernment, believes about eGovernment, decision on eGovernment adoption and eGovernment embracement

**3.2.1 ICT Accessibility.** Chart 1 finds out the intersection of urban/rural and internet access availability at home variables. The chart clearly shows the higher percentage of internet access at homes in urban (96.8) areas rather than those living in rural areas (89.5%).

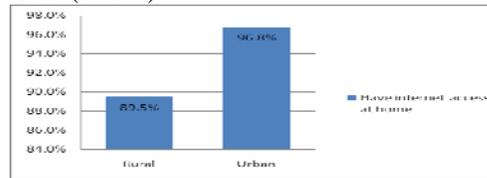

**Chart 1**: Internet accessibility within urban and rural areas

Another factor that will encourage eGovernment adoption is the quality of enabling ICT services within urban and rural areas in KSA. Chart 2 shows how satisfied the participants are with the ICT quality available at urban and rural areas. Most of the participants are fairly satisfied with the quality of the ICT services within urban and rural areas.

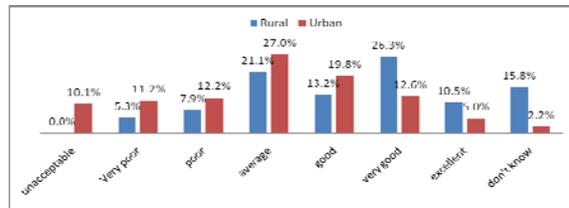

**Chart 2:** Perception of ICT quality in rural and urban areas.

There is a unimodal in Chart 3 for 'ICT cost' values except the last value (Don't know) which comprises 18.8% for rural areas and 2.2% for urban areas. The unimodal inclination in the middle of this chart illustrates that most participants believe that the costs of ICT services are between expensive and average.

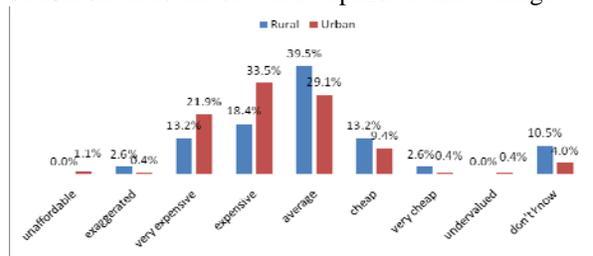

**Chart 3:** Perception of ICT costs in rural and urban areas.

It can be concluded that participants living in urban areas have more access to internet in their homes than those rural areas.

**3.2.2 ICT usage.** EGovernment adoption depends on internet usage since eGovernment transactions go through via ICT services. However, there are social factors that reduce the number of internet users such income, age and education resulting in reducing the number of eGovernment adopters.

Chart 4 represents the interaction of people who already experienced the internet with the age groups. There is a clear altitude in the percentage of people who experienced the internet between the age of 18 and 55, whereas there are fewer participants of who experienced the internet whom are older than fifty six. This indicates that age was not a major factor affecting the number of internet users for the participants.

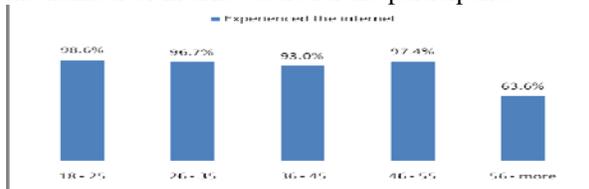

**Chart 4:** Percentage of people who experienced the internet intersected with the age groups percentage.

Chart 5 shows that participants between 18 and 45 years old had higher percentage of years or experience (more than 5 years). On the other hand, participants who were 56 older or older had less experience using the internet where the chart shows that 36.4% of them had less than one year of experience.

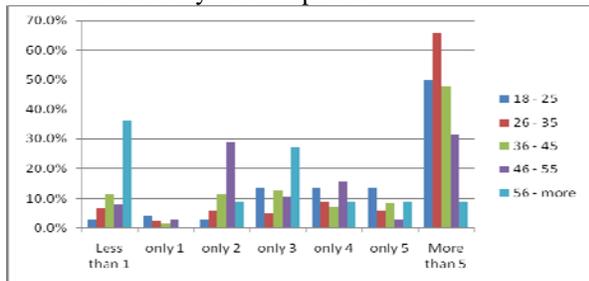

**Chart 5:** Years of experience percentage for the participants intersected with the age groups percentage.

Again Chart 6 the intersection between the participants who had experienced the internet and the education level shows that there is no relationship between education levels and internet adoption, since internet adoption is very high for all education levels. However, to be accurate this does not completely apply for the relationship between 'Technical qualification' and internet experience as other education level because of the slightly lower percentage (88.9%) which is relatively insignificant.

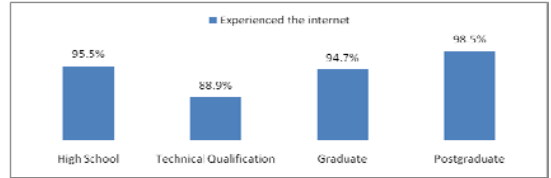

**Chart 6:** Percentage of people who experienced the internet intersected with the education level percentage.

Chart 7 illustrates the internet years of experience compared with the education. This chart shows that there is a direct relationship between higher levels of education and internet usage. However, participants who had high school certificates and technical qualifications had higher percentages in 2 and 3 years only.

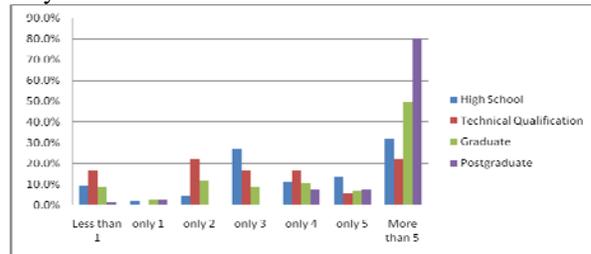

**Chart 7:** Years of experience percentage for the participants intersected with the education level percentage.

The relationship between total monthly incomes and internet years of experience in Chart 9 depicts the influence of higher income over internet years of experience. Incomes between 6001 and 1000 SR and those which were more than 10,000 SR had higher percentage of 'more than 5' years of internet experience. However, the difference between these incomes and other incomes are not very high in 'more than 5'.

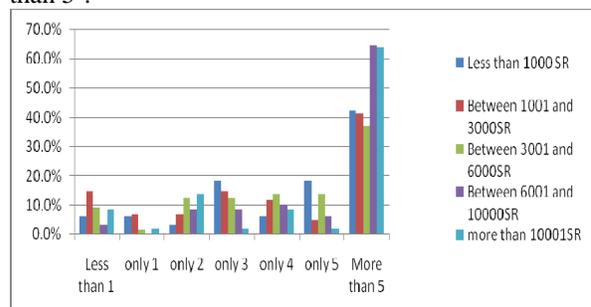

**Chart 9:** Years of experience percentage for the participants intersected with the total monthly income percentage.

There was not a very clear relationship between age, education and income and internet experience. However, there was a relationship between internet years of experience and these base variables, where when salary, education increases the years of experience increases while participants who where more than 56 had less experience in the internet.

**3.2.3 Perceptions about eGovernment.** Generally, perceptions about eGovernment can be affected by how it is advertised therefore, it is important to measure how participants think eGovernment can be advertised. Chart 10 shows participants think that: word of mouth and internet advertisements all together as the best method to increase eGovernment awareness.

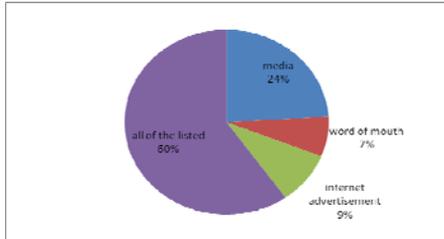

**Chart 10:** Methods perceived by participants as the best way to increase government awareness.

**3.2.4 Believes about eGovernment.** The nationality of eGovernment users and eCommerce previous experiences might affect on believes and trust of KSA eGovernment security and privacy. For example, participants might have had an experience of internet security privacy or privacy issue affecting believes about eGovernment.

Chart 15 and Chart 16 shows very similar trends towards participant's averagely acceptance of security and privacy levels of Saudi eGovernment. However, not acknowledging how secure or private eGovernment is very high with KSA visitors.

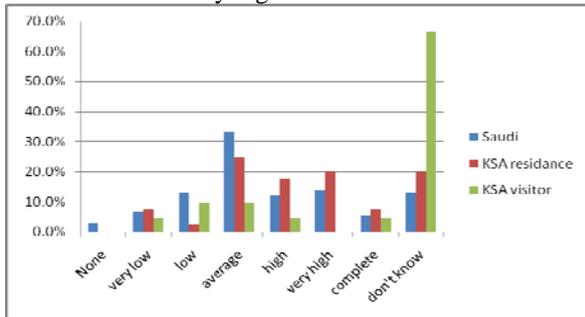

**Chart 15:** Nationality and security perception.

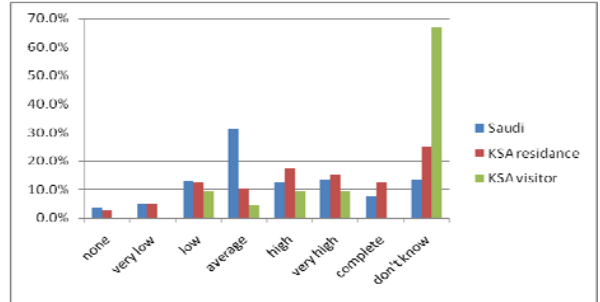

**Chart 16:** Nationality and privacy perception

Again compared to eCommerce experience; perceptions about eGovernment security and privacy were very similar as shown in Chart 17. Participants who experienced eCommerce had an average or very high trust in eGovernment security and privacy. This indicates a clear positive relationship between the users who experienced e-commence and believes about eGovernment security and privacy.

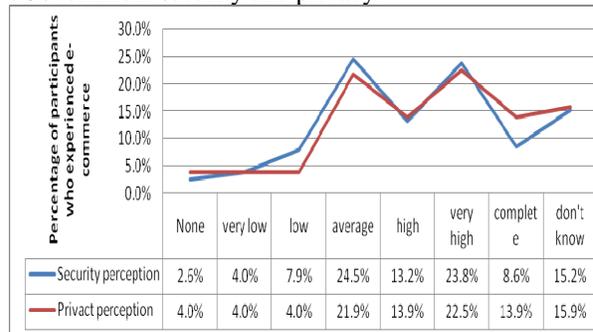

**Chart 17:** eCommerce experience intersecting with perception of eGovernment security and privacy.

Most of the participants had an average to very high trust in Saudi eGovernment security and privacy for different nationalities and for participants who experienced eCommerce. However, not acknowledging about eGovernment aspects is becoming more as a trend between the charts that is showing participants' opinions on Saudi eGovernment.

**3.2.5. Behavior intentions to adopt eGovernment.** Pervious disposition toward technology and learning about technology in education or training might have an influence on the perceptions of ease of use and usefulness of eGovernment especially when there are intentions to adopt eGovernment.

Chart 18 shows that most of participants 'don't know' about eGovernment ease of use and usefulness. Furthermore, high percentage of participants has an average perception for eGovernment ease of use and usefulness. However, there is a relatively high percentage of 9.1% for participants who believe that there is none perceived eGovernment usefulness.

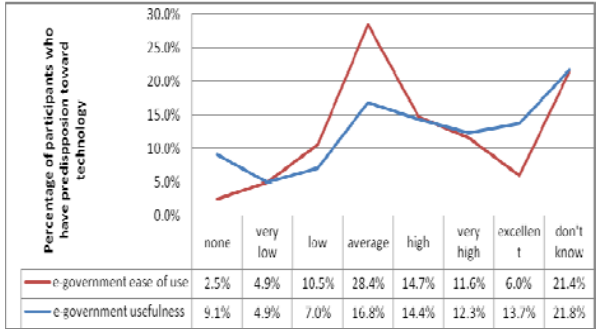

**Chart 18:** Predisposition toward technology intersecting with perception of eGovernment ease of use and usefulness.

Participants who trained or learned using technology in work-field or academic-field provided were more inclined towards average perception of eGovernment ease of use and usefulness as shown in Chart 19. Nevertheless, the highest percentage was to participants who actually 'don't know' about eGovernment ease of usefulness.

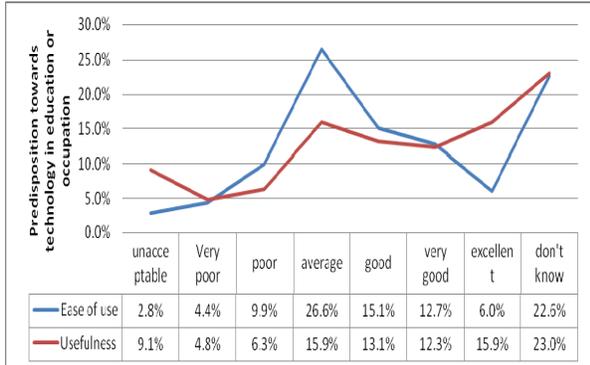

**Chart 19:** Predisposition toward technology in education or training intersecting with perception of eGovernment complexity and usefulness.

The percentage of participants who did not know about eGovernment ease of use or usefulness is relatively high. Additionally, this trend repeated for all previous variables related to eGovernment, raising an issue of eGovernment awareness.

### 3.2.6. Decision on eGovernment adoption.
Pervious eGovernment and eCommerce experiences might have an influence on how participants perceive the reliability which is a final determining factor of eGovernment after the adoption has been decided.

In Chart 20 participants who experienced eGovernment and eCommerce provided approximately identical opinions on eGovernment reliability. Most participants agree that Saudi eGovernment reliability is between very good and excellent.

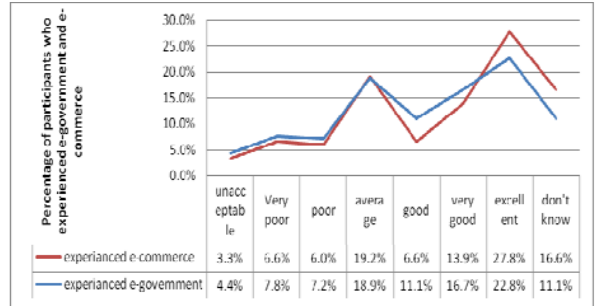

**Chart 20:** Intersection of reliability with participants who experienced eCommerce and eGovernment.

Most participants who experienced eCommerce or eGovernment considered Saudi eGovernment as a reliable media in conducting eGovernment services. However, it would be reasonable enough to say that people who didn't experience eGovernment and eCommerce won't actually know if eGovernment is reliable or not.

### 3.2.7. EGovernment Adoption and Implementation Related Issues.
There are issues that emerge after eGovernment is implemented by the government and adopted by the users such as: whether participants will consider eGovernment as the best method to contact the government and if they prefer to receive hardcopy evidence of their e-transaction status.

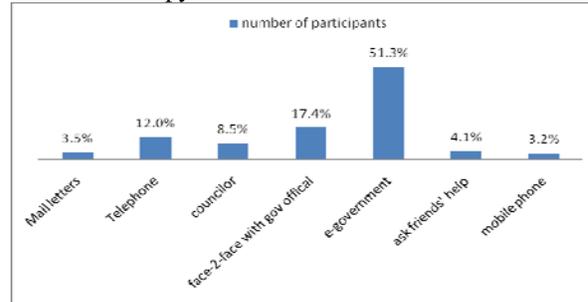

**Chart 21:** Perceived best method to contact the government.

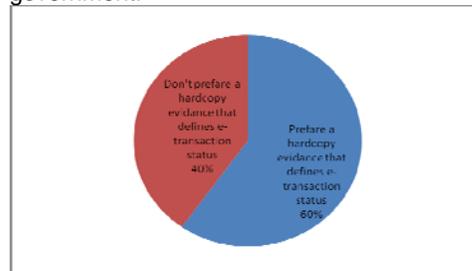

**Chart 22:** Best perceived method to increase eGovernment awareness

As Chart 21 and Chart 22 shows respectively that: most participants prefer to use 'eGovernment', prefer a hardcopy evidence of e-transaction status.

## 4. Conclusion

EGovernment success factors pertaining to the KSA, the focal point of this research, were introduced based on findings and data sources including: literature review and surveys of current and potential eGovernment users. The findings were collated, analyzed and then synthesized to identify factors that were considered pertinent to the successful adoption of eGovernment in the KSA.

Altogether, seven success factors were identified, namely: Improving ICT infrastructure in KSA, improving ICT services, improving technology literacy, planning and conducting a comprehensive eGovernment awareness programs, developing user-friendly, organized, well-supported eGovernment websites, provision of high quality and low fees services, provision of hardcopy of e-transactions' status.

There is little doubt that ICT infrastructure needs to improve in order to provide simple and fast access to eGovernment through the use of adequate interfaces including laptops, mobile phones, PDAs and Desktops. This improvement should be done with regards to the users' and eGovernment needs, therefore a comprehensive study inclusive of these needs is recommended.

The ICT usage uptake relies partially on the quality of ICT services. The improvement of these services should include lowering usage costs and increasing their quality. Still, these services require a level of supervision that guarantees the customers' right in to cheap and high quality services hopefully leading to strong eGovernment adoption.

The education sector should follow world's trends in new advancements embracing technology with a view to educate a younger generation who are easily able to adapt to eGovernment usage without any kind of training. Furthermore, the public sector should provide at least minimum technology training and education for its employees with a view to facilitate or improve on their use of computers and the internet to enable eGovernment adoption.

While technical and educational issues are of importance, getting the message across to the users is also important. Widespread information sessions should be conducted to guarantee that users everywhere within the KSA become aware of eGovernment implementation initiatives and become comfortable with the idea. It is proposed that the implementation of the information plan should be conducted via many channels including: conferences, news meetings with government officials, TV, radio, internet, and mobile phones, events (e.g. sports events) and word of mouth. Change agents can be used to promote eGovernment usage for people that are less inclined towards adoption by initially encouraging change agents to adopt eGovernment though incentives.

Skepticism also emerged as an issue. EGovernment users may be skeptic with regard to security and privacy and technical measures. These measures can protect the users and safeguard their information from any sabotaging or intruding acts.

The face of eGovernment will be reflected in the websites used, in other words they will be the face of the project. They should be well organized, simple to use, well maintained and supported. This would be an obvious sign of a project well on the way to successful implementation. The websites should be also developed in a way that enables equal opportunity for all users. In other words, the websites should consider users' needs and requirements including vision and hearing impairment, older generations and non-Arabic or non-English speakers.

The introduction of highly regarded eGovernment services would encourage eGovernment adoption. Furthermore, eGovernment fees should be kept low or free at least at the introduction of eGovernment akin to the eCommerce phenomena when products sold on the internet were cheaper than others.

Nevertheless, eGovernment adoption would be a transitional experience moving from paper to electronic means in conducting and acquiring government services. Therefore, it would need a change factor that helps users comprehend and trust eGovernment transactions. One way of achieving this trust may be via an option of sending a hardcopy of the electronic transaction status through snail-mail or fax to ease the user's concerns by providing hard evidence of the transaction process.

## 5 References

[1] Atkisson, A. (1999). "The innovation diffusion game", http://www.context.org/ICLIB/IC28/Atkisson.htm, Accessed on Mar 12, 2007.

[2] Abanumy, A. & Mayhew, P. 2005, 'M-government Implications For E-Government In Developing Countries: The Case Of Saudi Arabia', EURO mGOV 2005, Brighton, UK, pp. 1-6.

[3] AGIMO (Australian Government Information Management Office). (2006). "Barriers to e-government use", Commonwealth of Australia. http://www.agimo.gov.au/publications/2006/july/australians_use_of_and_satisfaction_with_e-

government_services__2006/barriers_to_e-government_use2, Accessed on June 11, 2007.

[4] Ajzen, I 1991, 'The theory of planned behavior, Organizational Behavior and Human Decision Processes, vol. 50, pp. 179-211.

[5] Al-Kibisi, G., de Boer, K., Mourshed, M. and Rea, N. (2001) Putting citizens on-line, not inline. The McKinsey Quarterly, Special Edition, pp. 64.

[6] Belanger, F & Hiller, JS 2006, 'A Framework for e-Government Implementation Privacy Implications', Business Process Management Journal, vol. 12, no. 1, pp. 48-60.

[7] Carter, L & Belanger, F 2003, 'The Influence of Perceived Characteristics of Innovating on e-Government Adoption', Electronic Journal of e-Government, vol. 2, no. 1, pp. 11-20.

[8] Carter, L & Belanger, F 2004, 'Citizen Adoption of Electronic Government Initiatives ', Proceedings of the 37th Hawaii International Conference on System Sciences, IEEE, Hawaii.

[9] Carter, L & Bélanger, F 2005, 'The utilization of e-government services: citizen trust, innovation and acceptance factors', Information System Journal, vol. 15, pp. 5-25.

[10] Chen, YN, Chen, HM, Huang, W & Ching, RKH 2006, 'E-Government Strategies in Developed and Developing Countries: An Implementation Framework and Case Study', Journal of Global Information Management, vol. 14, no. 1, pp. 23-47.

[11] Ebrahim, Z & Irani, Z 2005, 'E-government adoption: architecture and barriers', Business Process Management Journal, vol. 11, no. 5, pp. 589-611.

[12] Geray, O & Al Bastaki, MA 2005, 'Dubai eGovernment Initiative: Concept, Achievements and the Future Pillars of Success '. Dubai eGovernment.

[13] Ghaziri, H. (2003). Prerequisites for Building E-Government: The case of the Arab countries. In proceedings of the 2003 International Business Information Management Conference, December 16-18th, Cairo, Egypt.

[14] Gilbert, D, Balestrini, P and Littleboy, D 2004, 'Barriers and benefits in the adoption of e-government', The International Journal of Public Sector Management, vol. 17 no. 4, pp. 286-301.

[15] Gil-Garcia, JR and Pardo, TA 2006, 'Multi-Method Approaches to Digital Government Research: Value Lessons and Implementation Challenges', Proceedings of the 39th Annual Hawaii International Conference on System Sciences (HICSS'06) Track 4, IEEE.

[16] Holden, SH, Norris, DF, and Fletcher, PD 2003 'Electronic government at the local level: Progress to date and future issues', Public Performance and Management Review, Vol. 26, no. 4, pp. 325–44.

[17] Internet World Stats. (2007)." Saudi Arabia Internet Usage and Marketing Report", http://www.internetworldstats.com/me/sa.htm Accessed on April 1, 2007.

[18] Joshi, J. B. D., Ghafor, A., Aref, W. G., & Spafford, E. H. (2002). Security and privacy challenges of a digital government. In W. J. McIver, & A. K. Elmagarmid (Eds.), Advances in digital government. Technology, human factors, and policy. Norwell, MA7.

[19] Layne, K & Lee, J 2001, 'Developing fully functional E-government: A four stage model', Government Information Quarterly, vol. 18, no. 2, pp. 122-36.

[20] MCIT (Ministry of Communications and Information Technology). (2006). "E-government Action Plan", http://www.yesser.gov.sa/english/executive_plan.asp?id=item3, Accessed on Mar 30, 2007.

[21] Reffat, R. 2003, Developing a Successful E-Government, Working Paper, School of Architecture, Design Science and Planning, University of Sydney, Australia.

[22] Warkentin, M, Gefen, D, Pavlou, P and Rose, G 2002, 'Encouraging citizen adoption of e-government by building trust', Electronic Markets, vol. 12, pp. 157–62.